\documentclass[11pt]{revtex4}    
\usepackage{amssymb,epsf}
\usepackage{latexsym}
\begin{document}

\title{ Magnetic Strings in Einstein-Born-Infeld-Dilaton Gravity}
\author{M. H. Dehghani$^{1,2}$ \footnote{mhd@shirazu.ac.ir}, A. Sheykhi$^{3}$\footnote{sheykhi@mail.uk.ac.ir} and S. H.
Hendi$^{4}$\footnote{hendi@mail.yu.ac.ir}}

\address{$^1$ Physics Department and Biruni Observatory, Shiraz University, Shiraz 71454, Iran\\
         $^2$ Research Institute for Astrophysics and Astronomy of Maragha (RIAAM), Maragha, Iran\\
         $^3$ Department of Physics, Shahid Bahonar University, P.O. Box 76175-132, Kerman, Iran\\
         $^4$ Department of Physic, College of Science, Yasouj University, Yasouj 75914, Iran}

\begin{abstract}
A class of spinning magnetic string in $4$-dimensional
Einstein-dilaton gravity with Liouville type potential which produces a longitudinal nonlinear
electromagnetic field  is presented. These solutions have
no curvature singularity and no horizon, but have a conic geometry. In these
spacetimes, when the rotation parameter does not vanish, there exists an electric field, and therefore the
spinning string has a net electric charge which is proportional to the rotation parameter.
Although the asymptotic behavior of these solutions are neither
flat nor (A)dS, we calculate the conserved quantities of these
solutions by using the counterterm method.
We also generalize these four-dimensional solutions to the case of $(n+1)$-dimensional
rotating solutions with $k\leq[n/2]$ rotation parameters,
and calculate the conserved quantities and electric charge of them.
\end{abstract}

\maketitle
\section{Introduction\label{Intro}}
The Born-Infeld \cite{BI} type of generalizations of Abelian and
non-Abelian gauge theories have received a lot of interest in
recent years. This is due to the fact that such generalizations
appear naturally in the context of the superstring theory
\cite{Lei}. The nonlinearity of the electromagnetic field brings
remarkable properties to avoid the black hole singularity problem
which may contradict the strong version of the Penrose cosmic
censorship conjecture in some cases. Actually a new non-linear
electromagnetism was proposed, which produces a nonsingular exact
black hole solution satisfying the weak energy condition
\cite{Sal}, and has distinct properties from Bardeen black holes
\cite{Bar}. The Born-Infeld action including a dilaton and an
axion field, appears in the couplings of an open superstring and
an Abelian gauge field. This action, describing a
Born-Infeld-dilaton-axion system coupled to Einstein gravity, can
be considered as a non-linear extension of the Abelian field of
Einstein-Maxwell-dilaton-axion gravity. Exact static solutions of
Einstein-Born-Infeld (EBI) gravity in arbitrary dimensions with
positive, zero or negative constant curvature horizons have been
constructed \cite{Wil,Dey,Cai2}. Rotating solutions of Einstein
(Gauss-Bonnet)-Born-Infeld in various dimensions with flat
horizons have also been obtained \cite{DR,DH}. When a dilaton
field is coupled to gravity, it has profound consequences for the
black hole/string solutions. Many attempts have been done to
construct exact solutions of Einstein-Maxwell-dilaton (EMd) and
Einstein-Born-Infeld-dilaton (EBId) gravity. While exact static
dilaton black hole solutions of EMd gravity have been constructed
in \cite{CDB1, CDB2,MW,PW,CHM,Cai}, exact rotating black holes
solutions with curved horizons have been obtained only for some
limited values of the coupling constant\cite{Fr,kun,kunz}. For
general dilaton coupling, the properties of rotating charged
dilaton black holes only with infinitesimally small charge
\cite{Cas} or small angular momentum have been investigated
\cite{Hor2,Shey0, SR}. When the horizons are flat, rotating
solutions of EMd gravity with Liouville-type potential in four
\cite {DF} and $(n+1)$-dimensions have been constructed
\cite{SDRP}. The studies on the black hole solutions of EBId
gravity in three and four dimensions have been carried out in
\cite{YI} and \cite{yaz,SRM,Tor}, respectively. Thermodynamics of
$(n+1)$-dimensional EBId solutions with flat \cite{DHSR} and
curved horizons have also been explored \cite{Shey}. In the
scalar-tensor theories of gravity, black holes solution coupled to
a Born-Infeld nonlinear electrodynamics have also been studied
recently in \cite{yaz3}. The appearance of dilaton changes the
asymptotic behavior of the solutions to be neither asymptotically
flat nor (anti)-de Sitter [(A)dS]. There are two motivations for
exploring non asymptotically flat nor (A)dS solutions of Einstein
gravity. First, these solutions can shed some light on the
possible extensions of AdS/CFT correspondence. Indeed, it has been
speculated that the linear dilaton spacetimes, which arise as
near-horizon limits of dilatonic black holes, might exhibit
holography \cite{Ahar}. The second motivation comes from the fact
that such solutions may be used to extend the range of validity of
methods and tools originally developed for, and tested in the case
of, asymptotically flat or asymptotically AdS black holes.

On the other hand, there are many papers which are dealing directly with the
issue of spacetimes in the context of cosmic string theory \cite{Vil}. All
of these solutions are horizonless and have a conical geometry; they are
everywhere flat except at the location of the line source. An extension to
include the electromagnetic field has also been done \cite{Muk}.
Asymptotically AdS spacetimes generated by static and spinning magnetic
sources in three and four dimensional Einstein-Maxwell gravity with negative
cosmological constant have been investigated in \cite{Lem1,Lem2}. The
generalization of these rotating solutions to higher dimensions and higher
derivative gravity have also been done in \cite{Deh2} and \cite{Deh3},
respectively. In the context of electromagnetic cosmic string, it was shown
that there are cosmic strings, known as superconducting cosmic string, that
behave as superconductors and have interesting interactions with
astrophysical magnetic fields \cite{Wit}. The properties of these
superconducting cosmic strings have been investigated in \cite{Moss}.
Superconducting cosmic strings have also been studied in Brans-Dicke theory
\cite{Sen1}, and in dilaton gravity \cite{Fer}. Exact magnetic rotating
solutions in three dimensions have been considered in \cite{Dia} while, two
classes of magnetic rotating solutions in four and higher dimensional EMd
gravity with Liouville-type potential have been explored in \cite{Deh4} and
\cite{SDR}, respectively. These solutions are not black holes, and represent
spacetimes with conic singularities. In the absence of a dilaton field,
magnetic rotating solutions of $(n+1)$-dimensional EBI theory have also been
constructed \cite{Safar}.

Our aim in this paper is to construct $(n+1)$-dimensional horizonless
solutions of EBId gravity. The motivation for studying these kinds of
solutions is that they may be interpreted as cosmic strings. Cosmic strings
are topological defects that arise from the possible phase transitions in
the early universe, and may play an important role in the formation of
primordial structures. Besides there are two main reasons for studying
higher dimensional solutions of EBId gravity. The first originates from
string theory, which is a promising approach to quantum gravity. String
theory predicts that spacetime has more than four dimensions. For a while it
was thought that the extra spatial dimensions would be of the order of the
Planck scale, making a geometric description unreliable, but it has recently
been realized that there is a way to make the extra dimensions relatively
large and still be unobservable. This is if we live on a three dimensional
surface (brane) in a higher dimensional spacetime (bulk) \cite{RS,DGP}. In
such a scenario, all gravitational objects are higher dimensional. The
second reason for studying higher dimensional solutions has nothing to do
with string theory. Four dimensional solutions have a number of remarkable
properties. It is natural to ask whether these properties are general
features of the solutions or whether they crucially depend on the world
being four dimensional.

The outline of our paper is as follows: In section \ref{field}, we present
the basic field equations and general formalism of calculating the conserved
quantities. In section \ref{mag}, we obtain the magnetic rotating
solutions of Einstein equation in the presence of dilaton and nonlinear electromagnetic
fields, and explore their
properties. The last section is devoted to summary and conclusions.

\section{Field Equations and Conserved Quantities\label{field}}

We consider the $(n+1)$-dimensional action in which gravity is coupled to
dilaton and Born-Infeld fields with an action
\begin{eqnarray}
I_{G} &=&-\frac{1}{16\pi }\int_{\mathcal{M}}d^{n+1}x\sqrt{-g}\left( \mathcal{%
\ R}\text{ }-\frac{4}{n-1}(\nabla \Phi )^{2}-V(\Phi )+L(F,\Phi )\right)
\nonumber \\
&&-\frac{1}{8\pi }\int_{\partial \mathcal{M}}d^{n}x\sqrt{-h}\Theta (h),
\label{Act}
\end{eqnarray}
where $\mathcal{R}$ is the Ricci scalar curvature, $\Phi $ is the dilaton
field, $V(\Phi )$ is a potential for $\Phi $ and $F^{2}=F_{\mu \nu }F^{\mu
\nu }$ ( $F_{\mu \nu }=\partial _{\mu }A_{\nu }-\partial _{\nu }A_{\mu }$ is
the electromagnetic field tensor and $A_{\mu }$ is the electromagnetic
potential). The last term in Eq. (\ref{Act}) is the Gibbons-Hawking boundary
term which is chosen such that the variational principle is well-defined.
The manifold $\mathcal{M}$ has metric $g_{\mu \nu }$ and covariant
derivative $\nabla _{\mu }$. $\Theta $ is the trace of the extrinsic
curvature $\Theta ^{ab}$ of any boundary(ies) $\partial \mathcal{M}$ of the
manifold $\mathcal{M}$, with induced metric(s) $h_{ab}$. In this paper, we
consider the action (\ref{Act}) with a Liouville type potential,
\begin{equation}
V(\Phi )=2\Lambda e^{4\alpha \Phi /(n-1)},  \label{v1}
\end{equation}
where $\Lambda $ is a constant which may be referred to as the cosmological
constant, since in the absence of the dilaton field ($\Phi =0$) the action (%
\ref{Act}) reduces to the action of Einstein-Born-Infeld gravity with
cosmological constant \cite{Dey,Cai2}. The Born-Infeld, $L(F,\Phi )$, part
of the action is given by
\begin{equation}
L(F,\Phi ))=4\beta ^{2}e^{4\alpha \Phi /(n-1)}\left( 1-\sqrt{1+\frac{%
e^{-8\alpha \Phi /(n-1)}F^{2}}{2\beta ^{2}}}\right) .
\end{equation}
Here, $\alpha $ is a constant determining the strength of coupling of the
scalar and electromagnetic field and $\beta $ is called the Born-Infeld
parameter with dimension of mass. In the limit $\beta \rightarrow \infty $, $%
L(F,\Phi )$ reduces to the standard Maxwell field coupled to a dilaton field
\begin{equation}
L(F,\Phi )=-e^{-4\alpha \Phi /({n-1})}F^{2},
\end{equation}
and $L(F,\Phi )\rightarrow 0$ as $\beta \rightarrow 0$. It is convenient to
set
\begin{equation}
L(F,\Phi )=4\beta ^{2}e^{4\alpha \Phi /(n-1)}\mathcal{L}(Y).
\end{equation}
where
\begin{eqnarray}
\mathcal{L}(Y) &=&1-\sqrt{1+Y},  \label{LY} \\
Y &=&\frac{e^{-8\alpha \Phi /({n-1})}F^{2}}{2\beta ^{2}}.  \label{Y}
\end{eqnarray}
The equations of motion can be obtained by varying the action (\ref{Act})
with respect to the gravitational field $g_{\mu \nu }$, the dilaton field $%
\Phi $ and the gauge field $A_{\mu }$ which yields the following field
equations

\begin{eqnarray}
\mathcal{R}_{\mu \nu }&=&\frac{4}{n-1}\left( \partial _{\mu }\Phi \partial
_{\nu }\Phi +\frac{1}{4}g_{\mu \nu }V(\Phi )\right) -4e^{-4\alpha \Phi/({n-1}%
)}\partial _{Y}\mathcal{L}(Y)F_{\mu \eta }F_{\nu }^{\text{ }\eta }  \nonumber \\
&&+\frac{4\beta^2}{n-1} e^{4\alpha \Phi/({n-1})}\left[ 2Y\partial _{Y}%
\mathcal{L}(Y)-\mathcal{L}(Y)\right] g_{\mu \nu },  \label{FE1}
\end{eqnarray}
\begin{equation}
\nabla ^{2}\Phi =\frac{n-1}{8}\frac{\partial V}{\partial \Phi }+2 \alpha
\beta ^{2} e^{4\alpha \Phi/({n-1})}\left[ 2{\ Y}\partial _{Y}\mathcal{L}(Y)-%
\mathcal{L}(Y)\right] ,  \label{FE2}
\end{equation}
\begin{equation}
\nabla _{\mu }\left(e^{-4\alpha \Phi/({n-1})}\partial _{Y}\mathcal{L}%
(Y)F^{\mu \nu }\right) =0.  \label{FE3}
\end{equation}
In particular, in the case of the linear electrodynamics with $\mathcal{L}%
(Y)=-{\frac{1}{2}}Y$, the system of equations (\ref{FE1})-(\ref{FE3}) reduce
to the well-known equations of EMd gravity \cite{CHM}.

The conserved mass and angular momentum of the solutions of the
above field equations can be calculated through the use of the substraction
method of Brown and York \cite{BY}. Such a procedure causes the resulting
physical quantities to depend on the choice of reference background. For
asymptotically (A)dS solutions, the way that one deals with these
divergences is through the use of counterterm method inspired by (A)dS/CFT
correspondence \cite{Mal}. However, in the presence of a non-trivial dilaton
field, the spacetime may not behave as either dS ($\Lambda >0$) or AdS ($%
\Lambda <0$). In fact, it has been shown that with the exception
of a pure cosmological constant potential where $\alpha =0$, no
AdS or dS static spherically symmetric solution exist for
Liouville-type potential \cite{PW}. But, as in the case of
asymptotically AdS spacetimes, according to the domain-wall/QFT
(quantum field theory) correspondence \cite{Sken,RGC}, there may
be a suitable counterterm for the stress energy tensor which
removes the divergences. In this paper, we deal with the
spacetimes with zero curvature boundary [$R_{abcd}(h)=0$], and
therefore the counterterm for the stress energy tensor should be
proportional to $h^{ab}$. Thus, the finite stress-energy tensor in
$(n+1)$-dimensional Einstein-dilaton gravity with Liouville-type
potential may be written as \cite{RGC}
\begin{equation}
T^{ab}=\frac{1}{8\pi }\left[ \Theta ^{ab}-\Theta h^{ab}+\frac{n-1}{l_{%
\mathrm{eff}}}h^{ab}\right] ,  \label{Stres}
\end{equation}
where $l_{\mathrm{eff}}$ is given by
\begin{equation}
l_{\mathrm{eff}}^{2}=\frac{(n-1)(\alpha ^{2}-n)}{2\Lambda }e^{-4\alpha \Phi
/(n-1)}.  \label{leff}
\end{equation}
In the particular case $\alpha =0$, the effective $l_{\mathrm{eff}}^{2}$ of
Eq. (\ref{leff}) reduces to $l^{2}=-n(n-1)/2\Lambda $ of the AdS spacetimes.
The first two terms in Eq. (\ref{Stres}) is the variation of the action (\ref
{Act}) with respect to $h_{ab}$, and the last term is the counterterm which
removes the divergences. One may note that the counterterm has the same form
as in the case of asymptotically AdS solutions with zero curvature boundary,
where $l$ is replaced by $l_{\mathrm{eff}}$. To compute the conserved
charges of the spacetime, one should choose a spacelike surface $\mathcal{B}$
in $\partial \mathcal{M}$ with metric $\sigma _{ij}$, and write the boundary
metric in ADM (Arnowitt-Deser-Misner) form:
\begin{equation}
h_{ab}dx^{a}dx^{a}=-N^{2}dt^{2}+\sigma _{ij}\left( d\varphi
^{i}+V^{i}dt\right) \left( d\varphi ^{j}+V^{j}dt\right) ,
\end{equation}
where the coordinates $\varphi ^{i}$ are the angular variables
parameterizing the hypersurface of constant $r$ around the origin, and $N$
and $V^{i}$ are the lapse and shift functions, respectively. When there is a
Killing vector field $\mathcal{\xi }$ on the boundary, then the quasilocal
conserved quantities associated with the stress tensors of Eq. (\ref{Stres})
can be written as
\begin{equation}
Q(\mathcal{\xi )}=\int_{\mathcal{B}}d^{n-1}x\sqrt{\sigma }T_{ab}n^{a}%
\mathcal{\xi }^{b},  \label{charge}
\end{equation}
where $\sigma $ is the determinant of the metric $\sigma _{ij}$, $\mathcal{%
\xi }$ and $n^{a}$ are the Killing vector field and the unit
normal vector on the boundary $\mathcal{B}$. For boundaries with
timelike ($\xi =\partial /\partial t$) and rotational Killing
vector field ($\varsigma _{i}=\partial /\partial \phi ^{i}$), one
obtains the quasilocal mass and components of the total angular
momentum as
\begin{eqnarray}
M &=&\int_{\mathcal{B}}d^{n-1}x\sqrt{\sigma }T_{ab}n^{a}\xi ^{b},
\label{Mastot} \\
J_{i} &=&\int_{\mathcal{B}}d^{n-1}x\sqrt{\sigma
}T_{ab}n^{a}\varsigma _{i}^{b},  \label{Angtot}
\end{eqnarray}
Note that these quantities depend on the location of the boundary $\mathcal{B%
}$ in the spacetime, although each is independent of the particular choice
of foliation $\mathcal{B}$ within the surface $\partial \mathcal{M}$.

\section{Magnetic Rotating Solutions}\label{mag}

In this section we are going to obtain rotating horizonless
solutions of the field equations (\ref{FE1})-(\ref{FE3}). First,
we construct the rotating $4$-dimensional spacetimes generated by
a magnetic source which produces a longitudinal magnetic field.
Second, we generalize these $4$-dimensional solutions to the case
of $(n+1)$-dimensional solutions.

\subsection{Longitudinal magnetic field solutions\label{Lmag1}}

Here we want to obtain the $4$-dimensional solution of Eqs. (\ref{FE1})-(\ref
{FE3}) which produces a longitudinal magnetic fields along the $z$
direction. We assume the following form for the metric
\begin{eqnarray}
ds^{2} &=&-\frac{\rho ^{2}}{l^{2}}R^{2}(\rho )\left( \Xi dt-ad\phi \right)
^{2}+f(\rho )\left( \frac{a}{l}dt-\Xi ld\phi \right) ^{2} \nonumber \\
&&+\frac{d\rho ^{2}}{f(\rho )}+\frac{\rho ^{2}}{l^{2}}R^{2}(\rho
)dz^{2}, \label{metric1}
\end{eqnarray}
where $a$ is the rotation parameter and $\Xi =\sqrt{1+a^{2}/l^{2}}$. The
functions $f(\rho )$ and $R(\rho )$ should be determined and $l$ has the
dimension of length which is related to the cosmological constant $\Lambda $
for the case of Liouville-type potential with constant $\Phi $. The angular
coordinate $\phi $ is dimensionless as usual and ranges in $[0,2\pi ]$,
while $\rho $ and $z$ have dimension of length.

The electromagnetic field equation (\ref{FE3}) can be integrated immediately
to give
\begin{eqnarray}
F_{\phi \rho } &=&\frac{q\Xi le^{2\alpha \Phi }}{\left( \rho R\right) ^{2}%
\sqrt{1-\frac{q^{2}}{\beta ^{2}\left( \rho R\right) ^{4}}}},  \nonumber
\label{Ftr} \\
F_{t\rho } &=&-\frac{a}{\Xi l^{2}}F_{\phi \rho },
\end{eqnarray}
where $q$ is the charge parameter of the string. To solve the system of
equations (\ref{FE1}) and (\ref{FE2}) for three unknown functions $f(\rho )$%
, $R(\rho )$ and $\Phi (\rho )$, we make the ansatz
\begin{equation}
R(\rho )=e^{\alpha \Phi }.  \label{Rphi}
\end{equation}
Using (\ref{Rphi}), the electromagnetic fields (\ref{Ftr}) and the metric (%
\ref{metric1}), one can show that equations (\ref{FE1}) and (\ref{FE2}) have
solutions of the form
\begin{eqnarray}
f(\rho ) &=&\frac{\left( \Lambda -2\beta ^{2}\right) (\alpha
^{2}+1)^{2}b^{2\gamma }}{\alpha ^{2}-3}\rho ^{2(1-\gamma )}+\frac{m}{%
\rho ^{1-2\gamma}}  \label{f} \\
&&+2\beta ^{2}\left( \alpha ^{2}+1\right) b^{2\gamma }\rho
^{2\gamma -1}\int \rho ^{2(1-2\gamma)}\sqrt{\left( 1-\zeta\right)
}{d\rho },
\end{eqnarray}
\begin{equation}
\Phi (\rho )=\frac{\alpha }{1+\alpha ^{2}}\ln (\frac{b}{\rho }),  \label{phi}
\end{equation}
where $\gamma =\alpha ^{2}/(1+\alpha ^{2})$ and
\begin{equation}
\zeta \equiv \frac{q^{2}}{\beta ^{2}b^{4\gamma }\rho ^{4(1-\gamma
)}}. \label{eta}
\end{equation}
Equation (\ref{Ftr}) shows that $\rho $ should be greater than
$\rho _{0}=(q/\beta b^{2\gamma })^{1/(2-2\gamma)}$ in order to
have a real nonlinear electromagnetic field and consequently a
real spacetime. Indeed, as we will see below, we may remove the
region $\rho <\rho _{0}$ by a transformation. The integral can be
done in terms of hypergeometric function and can be written in a
compact form. The result is
\begin{eqnarray}
f(\rho ) &=&\frac{(\alpha ^{2}+1)^{2}b^{2\gamma }\rho ^{2(1-\gamma )}}{%
\alpha ^{2}-3}\left[ \Lambda +2\beta ^{2}\left( 1-\text{{\ }}%
_{2}F_{1}\left( \left[ -\frac{1}{2},\frac{\alpha ^{2}-3}{4}\right]
,\left[ \frac{\alpha ^{2}+1}{4}\right] ,\zeta \right) \right)
\right]   \nonumber
\\
&&+\frac{m}{\rho ^{1-2\gamma }},  \label{frhohyp}
\end{eqnarray}
where $b$ and $m$ are arbitrary constants. One may note that as
$\beta \longrightarrow \infty $ this solution reduces to the
$4$-dimensional
magnetic strings given in Ref. \cite{Deh4}. In the absence of dilaton field (%
$\alpha =\gamma =0$), the above solutions reduce to the $4$-dimensional
horizonless rotating solutions of Einstein-Born-Infeld gravity presented in
\cite{Safar}. One can easily show that the gauge potential $A_{\mu }$
corresponding to the electromagnetic tensor (\ref{Ftr}) can be written as
\begin{equation}
A_{\mu }=\frac{q}{\rho}\times \text{ }_{2}F_{1}\left( \left[ {\frac{1%
}{2},\frac{{\alpha }^{2}+1}{{4}}}\right] ,\left[ {\frac{{\alpha }^{2}{+5}}{{4%
}}}\right] ,\zeta \right) \left( \frac{a}{l}\delta _{\mu }^{t}-\Xi
l\delta _{\mu }^{\phi }\right).
\end{equation}
Now we study the properties of these solutions. To do this, we
first look for the curvature singularities in the presence of
dilaton field. It is easy to show that the Kretschmann invariant
$R_{\mu \nu \lambda \kappa }R^{\mu \nu \lambda \kappa }$ diverges
at $\rho =\rho _{0}$, it is finite for $\rho
>\rho _{0}$ and goes to zero as $\rho \rightarrow \infty $. Therefore one
might think that there is a curvature singularity located at $\rho
=\rho _{0} $. Two cases happen. In the first case the function
$f(\rho )$ has one or more real root(s) larger than $\rho _{0}$.
In this case the function $f(\rho )$ is negative for $\rho <r_{+}$%
, and positive for $\rho >r_{+}$  where $r_{+}$ is the largest
real root of $f(\rho )=0$. Indeed, $g_{\rho \rho }$ and $g_{\phi
\phi }$ are related by $f(\rho )=g_{\rho \rho }^{-1}=l^{-2}g_{\phi
\phi }$, and therefore when $g_{\rho \rho }$ becomes negative
(which occurs for $\rho _{0}<\rho <r_{+}$) so does $g_{\phi \phi
}$. This leads to an apparent change of signature of the metric from $+2$ to $%
+1$, and therefore indicates that we are using an incorrect extension. To
get rid of this incorrect extension, we introduce the new radial coordinate $%
r$ as
\begin{equation}
r^{2}=\rho ^{2}-r_{+}^{2}\Rightarrow d\rho ^{2}=\frac{r^{2}}{r^{2}+r_{+}^{2}}%
dr^{2}.
\end{equation}
With this new coordinate, the metric (\ref{metric1}) becomes
\begin{eqnarray}
ds^{2} &=&-\frac{b^{2\gamma }\left( r^{2}+r_{+}^{2}\right) ^{(1-\gamma )}}{%
l^{2}}\left( \Xi dt-ad\phi \right) ^{2}+f(r)\left( \frac{a}{l}dt-\Xi ld\phi
\right) ^{2}+\frac{r^{2}}{(r^{2}+r_{+}^{2})f(r)}dr^{2}  \nonumber \\
&&+\frac{b^{2\gamma }\left( r^{2}+r_{+}^{2}\right) ^{(1-\gamma )}}{l^{2}}%
dz^{2},  \label{metric2}
\end{eqnarray}
where the coordinates $r$ and $z$ assume the values $0\leq r<\infty $ and $%
-\infty \leq z<\infty $. The function $f(r)$ is now given as
\begin{eqnarray}
f(r) &=&\frac{(\alpha ^{2}+1)^{2}b^{2\gamma }}{\alpha ^{2}-3}%
(r^{2}+r_{+}^{2})^{(1-\gamma )}\left[\Lambda+2\beta ^{2}
\left( 1-\text{{\ }}_{2}F_{1}\left( \left[ -\frac{1}{2},\frac{\alpha ^{2}-3%
}{4}\right] ,\left[ \frac{\alpha ^{2}+1}{4}\right] ,\eta
\right) \right)\right]\nonumber \\
&&+\frac{m}{(r^{2}+r_{+}^{2})^{(1-2\gamma )/2}} , \label{f2}
\end{eqnarray}
where
\begin{equation}
\eta \equiv\frac{q^{2}}{\beta ^{2}b^{4\gamma
}(r^{2}+r_{+}^{2})^{2(1-\gamma )}}.
\end{equation}
The gauge potential in the new coordinate is
\begin{equation}
A_{\mu }=\frac{q}{(r^{2}+r_{+}^{2})^{1/2}}%
\times \text{ }_{2}F_{1}\left( \left[ {\frac{1}{2},\frac{{\alpha }^{2}+1}{{4}}}%
\right] ,\left[ \frac{\alpha^{2}+5}{4}\right] ,\eta \right)
\left( \frac{a}{l}\delta _{\mu }^{t}-\Xi l\delta _{\mu }^{\phi
}\right) .
\end{equation}
One can easily show that the Kretschmann scalar does not diverge in the
range $0\leq r<\infty $. However, the spacetime has a conic geometry and has
a conical singularity at $r=0$, since:
\begin{equation}
\lim_{r\rightarrow 0}\frac{1}{r}\sqrt{\frac{g_{\phi \phi }}{g_{rr}}}\neq 1
\label{limit}
\end{equation}
That is, as the radius $r$ tends to zero, the limit of the ratio
``circumference/radius'' is not $2\pi $ and therefore the spacetime has a
conical singularity at $r=0$. The canonical singularity can be removed if
one identifies the coordinate $\phi $ with the period
\begin{equation}
\text{Period}_{\phi }=2\pi \left( \lim_{r\rightarrow 0}\frac{1}{r}\sqrt{%
\frac{g_{\phi \phi }}{g_{rr}}}\right) ^{-1}=2\pi (1-4\mu ),  \label{period}
\end{equation}
where $\mu $ is given by
\begin{eqnarray}
\mu  &\equiv&\frac{1}{4}\left[ 1-\left( -\frac{\left( 2\,{\beta
}^{2}+\Lambda
\right) l{b}^{2\,\gamma }\left( {\alpha }^{2}+1\right) }{2}{r_{+}}%
^{1-2\,\gamma }+l{\beta }^{2}{b}^{2\,\gamma }\left( {\alpha }^{2}+1\right) \,%
{r_{+}}^{1-2\,\gamma }\right. \right.   \nonumber \\
&&\left. \left. \times \text{ }_{2}F_{1}\left( \left[ -\frac{1}{2},\frac{%
\alpha ^{2}-3}{4}\right] ,\left[ \frac{\alpha ^{2}+1}{4}\right]
,\eta
_{+0}\right) -\frac{2\,l{q}^{2}{b}^{-2\gamma }\left( {%
\alpha }^{2}+1\right) }{(\alpha ^{2}+1){r_{+}}^{(3-2\gamma
)}}\right.
\right.   \nonumber \\
&&\left. \left. \times \text{ }_{2}F_{1}\left( \left[ {\frac{1}{2},\frac{{%
\alpha }^{2}+1}{4}}\right] ,\left[ {\frac{{\alpha
}^{2}+5}{{4}}}\right] ,\eta _{+0}\right) \right) ^{-1}\right].
\label{mu}
\end{eqnarray}
Here we have defined $\eta _{+0}=\eta(r=0)$ and we have
expressed the mass parameter $m$ in terms of $q$ and $\Lambda $ by
using equation $f(r=0)=0$.
From Eqs. (\ref{limit})-(\ref{mu}), one concludes that near the origin $r=0$%
, the metric (\ref{metric2}) describes a spacetime which is
locally flat but has a conical singularity at $r=0$ with a deficit
angle $\delta \phi =8\pi \mu $. Since near the origin the metric
(\ref{metric2}) is identical to the spacetime generated by a
cosmic string, by use of Vilenkin procedure \cite {Vil2} $\mu $ of
Eq. (\ref{mu}) can be interpreted as the mass per unit length of
the string. In the second case, the function $f(\rho )$ is
positive for $\rho >\rho _{0}$, and the transformation $r^{2}=\rho
^{2}-\rho _{0}^{2}$ removes the imaginary part of the metric. In
this case, we have a spacetime with naked singularity which we are
not interested in it. In the rest of the paper we consider only
the first case.

Next we investigate the casual structure of the spacetime given in
Eq. (\ref {metric2}). The metric (\ref{metric2}) and the other
metric that we will present in this paper are neither
asymptotically flat nor (A)dS. As one can see from Eq. (\ref{f2}),
there is no solution for $\alpha =\sqrt{3}$. The cases with
$\alpha>\sqrt{3}$ and $\alpha <\sqrt{3}$ should be considered
separately. For $\alpha >\sqrt{3}$, as $r$ goes to infinity the
dominant
 term in Eq. (\ref{f2}) is the last term, and therefore the function $f(r)$ is positive
in the whole spacetime, despite the sign of the cosmological constant $%
\Lambda $, and is zero at $r=0$. Thus, the solution given by Eqs.
(\ref {metric2})-(\ref{f2}) exhibits a spacetime with conic
singularity at $r=0$. For $\alpha <\sqrt{3}$, the dominant term
for large values of $r$\ is the first term, and therefore the
function $f(r)$ given in Eq. (\ref{f2}) is positive in the whole
spacetime only for negative values of $\Lambda $. In this case the
solution describes a spacetime with conic singularity at $r=0$.
The solution is not acceptable for $\alpha <\sqrt{3}$ with
positive values of $\Lambda $, since the function $f(r)$ is
negative for large values of $r$.

Of course, one may ask for the completeness of the spacetime with $r\geq 0$
(or $\rho \geq r_{+}$) \cite{Lem2,Hor3}. It is easy to see that the
spacetime described by metric (\ref{metric2}) is both null and timelike
geodesically complete. In fact, we can show that every null or timelike
geodesic starting from an arbitrary point can either extend to infinite
values of the affine parameter along the geodesic or end on a singularity at
$r=0$. To do this, first, we perform the rotation boost
\begin{equation}
(\Xi t-a\phi )\mapsto t,\hspace{0.5cm}(at-\Xi l^{2}\phi )\mapsto l^{2}\phi ,
\end{equation}
in the $t-\phi $ plane. Then the metric (\ref{metric2}) becomes
\begin{eqnarray}
ds^{2} &=&-\frac{b^{2\gamma }\left( r^{2}+r_{+}^{2}\right) ^{(1-\gamma )}}{%
l^{2}}dt^{2}+\frac{r^{2}}{%
(r^{2}+r_{+}^{2})f(r)}dr^{2}+l^{2}f(r)d\phi ^{2} \\
&&+\frac{b^{2\gamma }\left( r^{2}+r_{+}^{2}\right) ^{(1-\gamma )}}{%
l^{2}}dz^{2}.
\end{eqnarray}
Using the geodesic equation, one obtains
\begin{eqnarray}
&&\dot{t}=\frac{l^{2}}{b^{2\gamma }(r^{2}+r_{+}^{2})^{1-\gamma
}}E, \hspace{0.7cm} \dot{\phi}=\frac{1}{l^{2}f(r)}L,\hspace{0.7cm}\dot{z}=\frac{l^{2}}{%
b^{2\gamma }(r^{2}+r_{+}^{2})^{1-\gamma }}P,  \label{Geo1} \\
&&r^{2}\dot{r}^{2}=(r^{2}+r_{+}^{2})f(r)\left[ \frac{l^{2}(E^{2}-P^{2})}{%
b^{2\gamma }(r^{2}+r_{+}^{2})^{1-\gamma }}-\xi \right] -\frac{r^{2}+r_{+}^{2}%
}{l^{2}}L^{2},  \label{Geo2}
\end{eqnarray}
where the dot denotes the derivative with respect to an affine parameter and
$\xi $ is zero for null geodesics and $+1$ for timelike geodesics. $E$, $%
L$ and $P$ are the conserved quantities associated with the
coordinates $t$, $\phi $ and $z$ respectively. Notice that $f(r)$
is always positive for $r>0$ and zero for $r=0$. First we consider
the null geodesics ($\xi =0$). (i) If $E^{2}>P^{2}$, the
spiraling particles ($L>0$) coming from infinity have a turning point at $%
r_{tp}>0$, while the nonspiraling particles ($L=0$) have a turning point at $%
r_{tp}=0$. (ii) If $E^{2}=P^{2}$ and $L=0$, then the velocities $\dot{r}$ and $%
\dot{\phi}$  vanish for any value of $r$, and therefore the null
particles moves on the $z$-axis. (iii) For $E^{2}=P^{2}$ and
$L\neq 0$, and also for $E^{2}<P^{2}$ and any values of $L$, there
is no possible null geodesic. Second, we analyze the timelike
geodesics ($\xi =+1$). Timelike geodesics are possible only if
$l^{2}(E^{2}-P^{2})>b^{2\gamma }r_{+}^{2(1-\gamma )}$.
In this case the turning points for the nonspiraling particles ($L=0$) are $%
r_{tp}^{1}=0$ and $r_{tp}^{2}$ given as
\begin{equation}
r_{tp}^{2}=\sqrt{[b^{-2\gamma }l^{2}(E^{2}-{P}^{2})]^{1/(1-\gamma
)}-r_{+}^{2}},
\end{equation}
while the spiraling ($L\neq 0$) timelike particles are bound
between $r_{tp}^{a}$ and $r_{tp}^{b}$ given by
\begin{equation}
0<r_{tp}^{a}\leq r_{tp}^{b}<r_{tp}^{2}.
\end{equation}
Therefore, we have confirmed that the spacetime described by Eq. (\ref
{metric2}) is both null and timelike geodesically complete.

Finally, we calculate the conserved quantities of these solutions.
The mass and angular momentum per unit length of the strings ($\alpha <\sqrt{3}$%
) can be calculated through the use of Eqs. (\ref{Mastot}) and (\ref{Angtot}%
). We find
\begin{equation}
{M}=\frac{b^{2\gamma }}{4}\left( \frac{(3-\alpha ^{2})\Xi ^{2}-2}{1+\alpha
^{2}}\right) m,
\end{equation}
\begin{equation}
J=\frac{b^{2\gamma }}{4}\left( \frac{3-\alpha ^{2}}{1+\alpha ^{2}}%
\right) \Xi m a.
\end{equation}
For $a=0$ ($\Xi =1$), the angular momentum per unit volume
vanishes, and therefore $a$ is the rotational parameter of the
spacetime. Of course, one may note that these conserved charges
are similar to the conserved charges of the magnetic rotating
dilaton string obtained in Ref. \cite{Deh4}. In the absence of
dilaton field ($\alpha =\gamma =0$) they reduce to that obtained
in \cite{Safar}.

\subsection{$(n+1)$-dimensional rotating solutions with all rotation
parameters\label{Lmag2}}

Our aim here is to construct the $(n+1)$-dimensional longitudinal magnetic
field solutions with a complete set of rotation parameters. The rotation
group in $n+1$ dimensions is $SO(n)$ and therefore the number of independent
rotation parameters is $[n/2]$, where $[x]$ denotes the integer part of $x$.
We now generalize the above solution given in Eq. (\ref{metric2}) with $%
k\leq \lbrack n/2]$ rotation parameters. This generalized solution can be
written as
\begin{eqnarray}
ds^{2} &=&-\frac{b^{2\gamma }\left( r^{2}+r_{+}^{2}\right) ^{(1-\gamma )}}{%
l^{2}}\left( \Xi dt-{{\sum_{i=1}^{k}}}a_{i}d\phi ^{i}\right) ^{2}+f(r)\left(
\sqrt{\Xi ^{2}-1}dt-\frac{\Xi }{\sqrt{\Xi ^{2}-1}}{{\sum_{i=1}^{k}}}%
a_{i}d\phi ^{i}\right) ^{2}  \nonumber
\\
&&+b^{2\gamma }\left( r^{2}+r_{+}^{2}\right) ^{(1-\gamma )}{{\sum_{i=1}^{%
n-k-2}}}(d\psi ^{i})^{2}+\frac{b^{2\gamma }(r^{2}+r_{+}^{2})^{(1-\gamma )}}{l^{2}(\Xi ^{2}-1)}%
{\sum_{i<j}^{k}}(a_{i}d\phi _{j}-a_{j}d\phi _{i})^{2}\nonumber \\
&&+\frac{r^{2}dr^{2}}{(r^{2}+r_{+}^{2})f(r)}+\frac{%
b^{2\gamma }\left( r^{2}+r_{+}^{2}\right) ^{(1-\gamma )}}{l^{2}}%
dz^{2},  \label{metric3}
\end{eqnarray}
where $\Xi =\sqrt{1+\sum_{i}^{k}a_{i}^{2}/l^{2}}$ and $f(r)$ is given as
\begin{eqnarray}
f(r) &=&\frac{2(\alpha ^{2}+1)^{2}b^{2\gamma
}(r^{2}+r_{+}^{2})^{(1-\gamma )}}{
(n-1)(\alpha ^{2}-n)}\left[ \Lambda +2\beta ^{2}\left( 1-\text{{\ }}%
_{2}F_{1}\left( \left[ -\frac{1}{2},\frac{\alpha
^{2}-n}{2n-2}\right] ,\left[ \frac{\alpha ^{2}+n-2}{2n-2}\right]
,\eta\right) \right) \right]   \nonumber \\
&&+\frac{m}{(r^{2}+r_{+}^{2})^{\left[ (n-1)(1-\gamma )-1\right]
/2}}.
\end{eqnarray}
The gauge potential is
\begin{eqnarray}
A_{\mu } &=&\frac{qb^{(3-n)\gamma }}{\lambda (r^{2}+r_{+}^{2})^{\lambda /2}}%
\text{ }_{2}F_{1}\left( \left[ {\frac{1}{2},\frac{{n+\alpha }^{2}{-2}}{{%
2(n-1)}}}\right] ,\left[ {\frac{{3n+\alpha }^{2}{-4}}{{2(n-1)}}}\right]
,\eta\right)    \nonumber  \label{A3} \\
&& \times \left( \sqrt{\Xi ^{2}-1}\delta _{\mu }^{t}-\frac{\Xi }{\sqrt{\Xi ^{2}-1}}%
a_{i}\delta _{\mu }^{i}\right) \hspace{0.5cm}{\text{(no sum on
i)}},
\end{eqnarray}
where $\lambda =(n-3)(1-\gamma )+1$. Again this spacetime has no
horizon and curvature singularity. However, it has a conical
singularity at $r=0$. One should note that these solutions
reduce to those discussed in \cite{Safar}, in the absence of dilaton field ($%
\alpha =\gamma =0$) and those presented in \cite{SDR} as $\beta \rightarrow
\infty $.

Next we calculate the conserved quantities of the
$(n+1)$-dimensional solutions. The mass and
angular momentum per unit length of the strings ($\alpha <\sqrt{n}$%
) can be calculated through the use of Eqs. (\ref{Mastot}) and (\ref{Angtot}%
). We find
\begin{equation}
{M}=\frac{(2\pi )^{n-3}b^{(n-1)\gamma }}{4}\left( \frac{(n-\alpha ^{2})\Xi
^{2}-(n-1)}{1+\alpha ^{2}}\right) m,  \label{M1}
\end{equation}
\begin{equation}
J_{i}=\frac{(2\pi )^{n-3}b^{(n-1)\gamma }}{4}\left( \frac{n-\alpha ^{2}}{%
1+\alpha ^{2}}\right) \Xi ma_{i}.  \label{J}
\end{equation}
For $a_{i}=0$ ($\Xi =1$), the angular momentum per unit volume
vanishes, and therefore $a_{i}$'s are the rotational parameters of
the spacetime. One may note that these conserved quantities are
similar to the conserved quantities of the $(n+1)$-dimensional
magnetic rotating dilaton string obtained in Ref. \cite{SDR}.

Finally, we calculate the electric charge of the solutions
(\ref{metric2}) and (\ref{metric3}) obtained in this section. To
determine the electric field we should consider the projections of
the electromagnetic field tensor on special hypersurfaces. The
normal to such hypersurfaces is
\begin{equation}
u^{0}=\frac{1}{N},\text{ \ }u^{r}=0,\text{ \ }u^{i}=-\frac{V^{i}}{N},
\end{equation}
and the electric field is $E^{\mu }=g^{\mu \rho }e^{-4\alpha \phi
/(n-1)}F_{\rho \nu }u^{\nu }$. Then the electric charge per unit length can
be found by calculating the flux of the electric field at infinity, yielding
\begin{equation}
{Q}=\frac{(2\pi )^{n-3}}{4}\sqrt{\Xi ^{2}-1}q.  \label{chden}
\end{equation}
It is worth noticing that the electric charge of the system per unit volume
is proportional to the rotation parameter, and is zero for the case of a
static solution. Again, in the absence of a non-trivial dilaton ($\alpha
=\gamma =0$), these conserved charges reduce to the conserved charges of $%
(n+1)$-dimensional horizonless rotating solutions of
Einstein-Born-Infeld gravity presented in \cite{Safar}.

\section{Summary and Conclusions}
To sum up, the Einstein-Born-Infeld action including a dilaton
field, appears in the couplings of an open superstring and an
Abelian gauge field. This action can be considered as a non-linear
extension of the Abelian field of Einstein-Maxwell-dilaton theory.
It is worth finding exact solutions of
Einstein-Born-Infeld-dilaton gravity for an arbitrary value of the
dilaton coupling constant, and investigate how the properties of
the solutions are modified when a dilaton is present. In this
paper, we constructed a new analytic solution of the
$4$-dimensional Einstein-Born-Infeld-dilaton theory in the
presence of Liouville-type potential for the dilaton field. These
solutions describe $4$-dimensional rotating strings with a
Longitudinal magnetic field and they have conic singularity at
$r=0$. Besides, they are horizonless without curvature
singularity. Because of the presence of the dilaton field, these
solutions are neither asymptotically flat nor (A)dS. In the
presence of Liouville-type potential, we obtained exact solutions
provided $\alpha \neq \sqrt{3}$. In the absence of a dilaton field
($\alpha =\gamma =0$), these solutions reduce to the
$4$-dimensional horizonless rotating solutions of
Einstein-Born-Infeld gravity presented in \cite{Safar}, while in
the case $\beta \rightarrow \infty $ they reduce to the
$4$-dimensional magnetic rotating dilaton string given in Ref.
\cite{Deh4}. We confirmed that these solutions are both null and
timelike geodesically complete by showing that in this spacetime
every null or timelike geodesic starting from an arbitrary point
can either extend to infinite values of the affine parameter along
the geodesic or end on a singularity at $r=0$. For the rotating
string, when the rotation parameter is nonzero, the string has a
net electric charge density which is proportional to the magnitude
of the rotation parameter. We also generalized these
four-dimensional solutions to the case of $(n+1)$-dimensional
magnetic rotating solutions with $k\leq[n/2]$ rotation parameters.
Finally, we calculated the conserved quantities of these solutions
by using the counterterm method inspired by the AdS/CFT
correspondence.

\acknowledgments{This work was partially supported by Research
Institute for Astrophysics and Astronomy of Maragha, Iran, and
also by Shahid Bahonar University of Kerman.}

\end{document}